\documentclass[doublecol]{epl2}
\usepackage{graphicx}%
\usepackage{psfrag}
\usepackage{dcolumn}
\usepackage{amsmath}
\usepackage{amssymb}
\usepackage{latexsym}

\def\bea{\begin{eqnarray}}
\def\eea{\end{eqnarray}}
\def\beq{\begin{equation}}
\def\eeq{\end{equation}}
\def\f{\frac}
\def\k{\kappa}

\def\be{\beta}

\def\t{\tau}
\def\r{\rho}
\def\a{\alpha}

\def\la{\langle}
\def\ra{\rangle}
\def\nn{\nonumber}

\def\l{\lambda}

\def\g{\gamma}

\def\a{\alpha}

\def\r{\rho}

\def\la{\langle}
\def\ra{\rangle}

\def\g{\gamma}

\def\f{\frac}

\title{Dynamics of path aggregation in the presence of turnover}
\author{Debasish Chaudhuri\inst{1} \and Peter Borowski\inst{2} \and
P. K. Mohanty\inst{3} \and Martin Zapotocky\inst{1,4}}
\institute{
\inst{1}
Max Planck Institute for the Physics of Complex Systems, 
N{\"o}thnitzer Strasse 38, 
01187 Dresden, Germany

\inst{2}
Department of Mathematics,
University of British Columbia, Vancouver, BC, Canada V6T 1Z2

\inst{3}
TCMP Division, Saha Institute of Nuclear Physics, 1/AF Bidhan Nagar,
Kolkata 700064, India

\inst{4}
Institute of Physiology,
Academy of Sciences of the Czech Republic,
Videnska 1083,
14220 Praha 4,
Czech Republic
}

\date{\today}

\abstract{
We investigate the slow time scales that arise 
from  aging of the paths during the process of path aggregation.
This is studied using Monte-Carlo simulations of a model
aiming to describe the formation of 
fascicles of axons mediated by contact axon-axon interactions.  The growing 
axons are represented as interacting directed random walks in two spatial 
dimensions. 
To mimic axonal turnover, random walkers are injected and whole 
paths of individual walkers are removed at specified rates. 
We identify several distinct time scales that emerge from the 
system dynamics and can exceed the average axonal lifetime by 
orders of magnitude.
In the dynamical steady state, the position-dependent distribution 
of fascicle sizes obeys a scaling law. 
We discuss our findings in terms of an analytically tractable, 
effective model of fascicle dynamics.
}

\begin{document}
\maketitle
\section{Introduction}
The process of path aggregation
is a ubiquitous phenomenon in nature. Some examples of such
phenomena are  river basin formation\cite{scheidegger},
aggregation of trails of liquid droplets moving down a window pane,
formation of insect pheromone trails\cite{ant1, ant2, pheromone},
and of pedestrian trail systems\cite{human,helbing}.


Path aggregation has been mathematically studied mainly in two classes 
of models.  One of them is known as the active-walker 
models\cite{helbing} 
in which each walker in course of its passage through the system
changes the surrounding environment locally, which in turn influences the later 
walkers. An example of such a process is the ant trail 
formation\cite{ant1,ant2}.
While walking, an ant leaves a chemical trail of pheromones which other
ants can sense and follow. 
The mechanism of human and animal trail formations is mediated by the
deformation of vegetation that generates an interaction between 
earlier and later walkers\cite{helbing}. A mathematical formalism to
study the formation of
such trails has been developed in Ref.\cite{human,helbing}.
The other class of models showing path aggregation deals with non-interacting
random walkers moving through a fluctuating environment.
In Ref.\cite{wilk}, condensation of trails of particles moving in an
environment with Gaussian spatial and temporal correlation is 
demonstrated analytically.  
Another  example of this model class is 
the Scheidegger river model\cite{scheidegger} 
(and related models\cite{river-book})
which describes the formation of a stream network by aggregation of streams 
flowing downhill on a slope with local random elevations.

In this Letter we analyze the dynamics of path aggregation using
a simple model that belongs to the class of active walker systems 
discussed above.
The model is similar to the one used to study path localization
in Ref.\cite{schulz}. In our model, however, we take into account the
aging of the paths, an important aspect of the active walker models. 
For instance,
in ant trail systems, the pheromone trails age due to evaporation. 
In the mammalian trail formation the deformations of the 
vegetation due to the movement of a mammal  
decays continuously with time\cite{helbing}.
In our model, the individual paths do not age gradually, but rather 
maintain their full identity until they are abruptly removed from 
the system. This particular rule for path aging is chosen to allow
application of our model to the process of axon fasciculation, which
we discuss next. 

During the development of an organism, 
neurons located at peripheral tissues 
(e.g. the retina or the olfactory epithelium)
establish connections to the brain via growing axons.
The {\it growth cone} structure at the tip of the axon interacts with 
other axons or external chemical signals  and can 
bias the direction of growth when spatially distributed chemical signals are 
present \cite{Gilbert}.  In the absence of directional 
signals, the growth cone maintains an approximately constant average 
growth direction, while exploring stochastically the environment in the 
transverse direction \cite{katz1985}. 
The interaction of growth cones with the shafts of other axons commonly 
leads to fasciculation of axon shafts\cite{Mombaerts}.
During development a significant portion of fully grown neurons die and
get replaced by newborn neurons with newly growing axons. 
For certain types of neurons (such as the sensory neurons of the mammalian
olfactory system) the turnover persists throughout the lifespan 
of an animal.
In mice,  the average lifetime of an olfactory 
sensory neuron is 1--2 months \cite{Crews}, which is less 
than one tenth of the mouse lifespan. 
The mature connectivity pattern is fully established only after several 
turnover periods \cite{Zou04}. 

The model we propose in this Letter captures the basic ingredients
of the process of axon fasciculation, i.e. attractive interaction of
growth cones with axon shafts, as well as neuronal turnover.
The main contribution of this Letter is a detailed discussion of the slow 
time scales that emerge from the dynamics of our model. 
Using Monte-Carlo simulations we characterize the time scale for the approach
to steady state and the correlation time within the steady state, and show 
that they can exceed the average axonal life time by orders of magnitude.
To understand these results we formulate
an analytically tractable effective single fascicle dynamics. 
This allows us to relate the observed slow time scales to the dynamics of
the basins of the fascicles.
From the effective fascicle dynamics we derive three time scales which
we compare to the time scales extracted from the Monte-Carlo simulation
of the full system.

For clarity, we stress that the dynamics of our model differs
substantially from  one-dimensional coalescence 
($A + A \to A$)\cite{benAvraham} or 
aggregation ($mA + nA \to (m+n)A$)\cite{redner}. 
In  our model, there is no direct
inter-walker interaction; rather, each random walker interacts locally 
with the {\em trails} of other walkers.
While the stationary
 properties of the system (such as the fascicle size distribution in the 
steady state) may be approximately understood using an analogy to 
one-dimensional diffusion with aggregation, the dynamical properties 
(such as the time scale of approach to the steady state, 
and the correlation time in the steady state) 
are undefined in the one dimensional analogy, and require understanding
 based on the full two dimensional model.
This is in contrast to the situation for  path aggregation 
models without turnover: e.g., the Scheidegger river 
network model can be mapped onto the Takayasu model of
diffusion-aggregation in presence of injection of mass
in one dimension\cite{takayasu}. 

\section{Model and numerical implementation}
\begin{figure}[t]
\psfrag{x}{$x$}
\psfrag{y}{$y$}
\psfrag{(a)}{$(a)$}
\includegraphics[width=4cm]{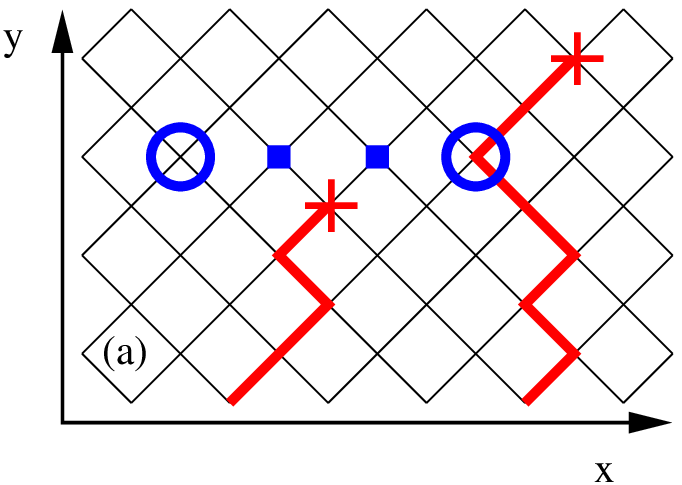}
\psfrag{Lb}{$D$}
\psfrag{Eb}{$E$}
\psfrag{y}{$y$}
\psfrag{(b)}{($b$)}
\includegraphics[width=4cm]{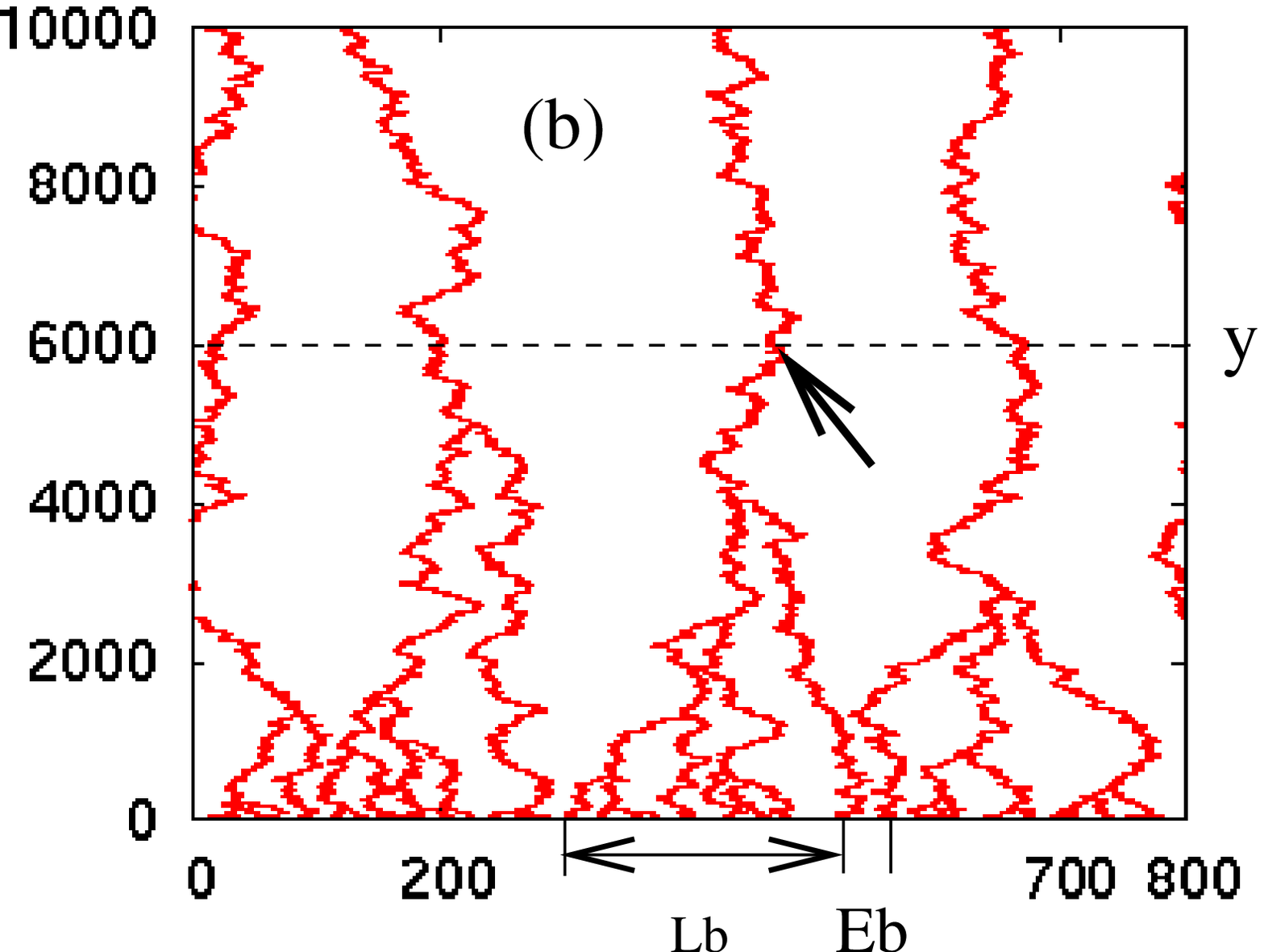} 
\caption{(Color online)
($a$)~Interacting directed random walks on a tilted square lattice. 
A random walker
($+$) represents a growth cone. For one walker, the possible future sites 
($\Box$) and their nearest neighbours ($\circ$) are marked. The trail of 
a walker (line) models an axon shaft.
($b$)~Typical late-time configuration ($t=25T$) in a system with 
$L=800$ and $N_0=100$. 
For the fascicle identified at  $y=6000$ (arrow), $D$ indicates its
basin, i.e. the interval at the level $y=0$ between the right-most and 
left-most axons belonging to the fascicle. The gap $E$ is the inter-basin free
space at $y=0$.
Note that the $y$-coordinate cannot be understood as equivalent to time
(see the main text).
}
\label{conf}
\end{figure}

Each growing axon is represented as a directed random walk in two spatial 
dimensions (Fig.~\ref{conf}$a$). The random walkers (representing the growth 
cones) are initiated at the periphery ($y=0$, random $x$) with a birth rate 
$\a$, and move towards the target area (large $y$)
with constant velocity $v_y=1$. In 
the numerical implementation on a tilted square lattice, at each time step 
the growth cone at ($x,y$) can move to ($x-1,y+1$) (left) or ($x+1,y+1$) 
(right). (Note that the sites are labeled alternatively by even $x$ and
odd $x$ at successive $y$ levels.)
The probability $p_{\{L,R\}}$ to move left/right is evaluated based 
on the axon occupancy at the ($x-1,y+1$) and ($x+1,y+1$) sites and their 
nearest neighbours (see Fig.~\ref{conf}$a$). In the simplest version of the 
model, the interaction is governed 
by the ``always attach, never detach'' rule: $p_L = 1$ when among the sites  
$(x \pm 1,y+1)$,  $(x \pm 3,y+1)$ only  $(x -3,y+1)$ is already occupied; 
$p_R =1$ when only  $(x+3,y+1)$ is occupied; $p_L = p_R = 1/2$ in all other 
cases.  
Periodic boundary conditions are used in the $x$-direction. 

To capture the effect of neuronal turnover, each random walker is assigned 
a lifetime from an exponential distribution with mean $T$. When the lifetime 
expires, the random walker and its entire trail is removed from the system. 
The mean number of axons in the system therefore reaches the steady state 
value $N=N_0\exp(-\be y)$ where
$N_0=\a/\be$, and $\beta = 1/T$ is the death rate per axon.  
In the simulations, we use $T = 10^{5}$ time steps, and restrict 
our attention to $y \le T/10$. The birth rate $\alpha$ is 
chosen so as to obtain the desired number of axons $N_0$, or equivalently, the 
desired axon density $\rho = N_0/L$ ($\r=1/2$ implies an average 
occupancy of one axon per site), where $L$ is the system size in 
$x$-direction. The presence of turnover distinguishes our model from
previous theoretical works on axon fasciculation \cite{AvanOoyen1,AvanOoyen2}.
 The $y$-coordinate in 
Fig.\ref{conf}$a$ cannot be
viewed as equivalent to time, and the dynamics at fixed $y$ (which
is the main focus of this Letter) has no analogy in one dimensional models of
aggregation or coalescence.

{\em Mean fascicle size:} 
A typical late-time configuration for a system with $L=800$ and $N_0=100$
is shown in Fig.~\ref{conf}$b$. With increasing fasciculation distance $y$, 
the axons  aggregate into a decreasing number $m(y)$ of {\em fascicles}. 
(At a given $y$, two axons are considered to be part of the same fascicle 
if they are not separated by any unoccupied sites.) The number of axons in 
the fascicle is referred to as the fascicle size $n$. The mean 
fascicle size $\bar n$ at level $y$ may be estimated using the following 
mean-field argument. Each of the $m$ fascicles collects axons that 
were initiated on an interval of length $D \simeq L/m = L\bar n/N$ at the 
level $y=0$ (see  Fig.~\ref{conf}$b$). The axons initiated at opposite edges of 
the interval are expected to meet within $y\simeq (D/2)^2$ steps of the 
random walk in $x$-direction. Consequently, 
$\bar n\simeq DN/L \simeq 2 \r y^{1/2} \exp(-\be y)$ 
for $y$ up to $y \simeq (L/2)^2$, 
where complete fasciculation ($\bar n=N$) is expected.
Thus for $y\ll (L/2)^2$ and $\be y\ll 1$ (which are satisfied in our 
simulations) one obtains the power law growth $\bar n \simeq 2\r y^{1/2}$.

\begin{figure}[t]
\begin{center}
\psfrag{t/T}{$t/T$}
 \psfrag{ninf-n}{$n_\infty - \la \bar n \ra $}
 \psfrag{ninf-c}{$n_\infty-c$}
 \psfrag{1y0.48}{\tiny{$~~~~~~y^{0.48}$}}
 \psfrag{0.25y0.48}{\tiny{$0.25\, y^{0.48}$}}
 \psfrag{y}{$y$}
 \psfrag{ y=102x}{\tiny{$y=10^2$}}
 \psfrag{ y=103x}{\tiny{$y=10^3$}}
 \psfrag{ y=104x}{\tiny{$y=10^4$}}
 \psfrag{ 5000*x}{\tiny{$\r=1/8$}}
\includegraphics[width=6cm]{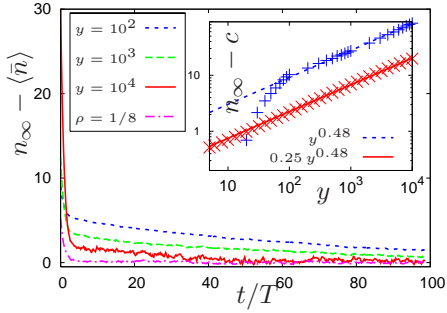}
\end{center}
\caption{(Color online)
Approach to the steady state in the $L=400$, $N_0=200$ system 
at representative $y$-levels indicated in the legend. 
The mean fascicle size $\la \bar n (t;y)\ra $ 
[averaged over $10^3$ initial conditions] 
approaches $n_\infty(y)$ as $t\to\infty$. 
The data set labeled as $\r=1/8$ is from the $L=400$, $N_0=50$
system, collected at $y=5000$.
Inset: Power-law growth of mean fascicle size 
$n_\infty - c = 2\r y^{b}$ in the steady state. 
The values of the growth exponent and offset are:  
$b=0.48$, $c=1.3$ 
for the $L=800$, $N_0=100$ system($\times$), and 
$b=0.48$, $c=6.9$ for the $L=400$, $N_0=200$ system($+$).}
\label{ap}
\end{figure}

\section{Slow time scales}
\begin{figure}[t]
\psfrag{c(t)}{$g(t)$}
 \psfrag{t/T}{$t/T$}
 \psfrag{(a)}{$(a)$}
 \psfrag{y=10xxxx}{\tiny{$y=10~~$}}
 \psfrag{y=100xxx}{\tiny{$y=10^2~~$}}
 \psfrag{y=1000xx}{\tiny{$y=10^3~~$}}
\includegraphics[width=4cm]{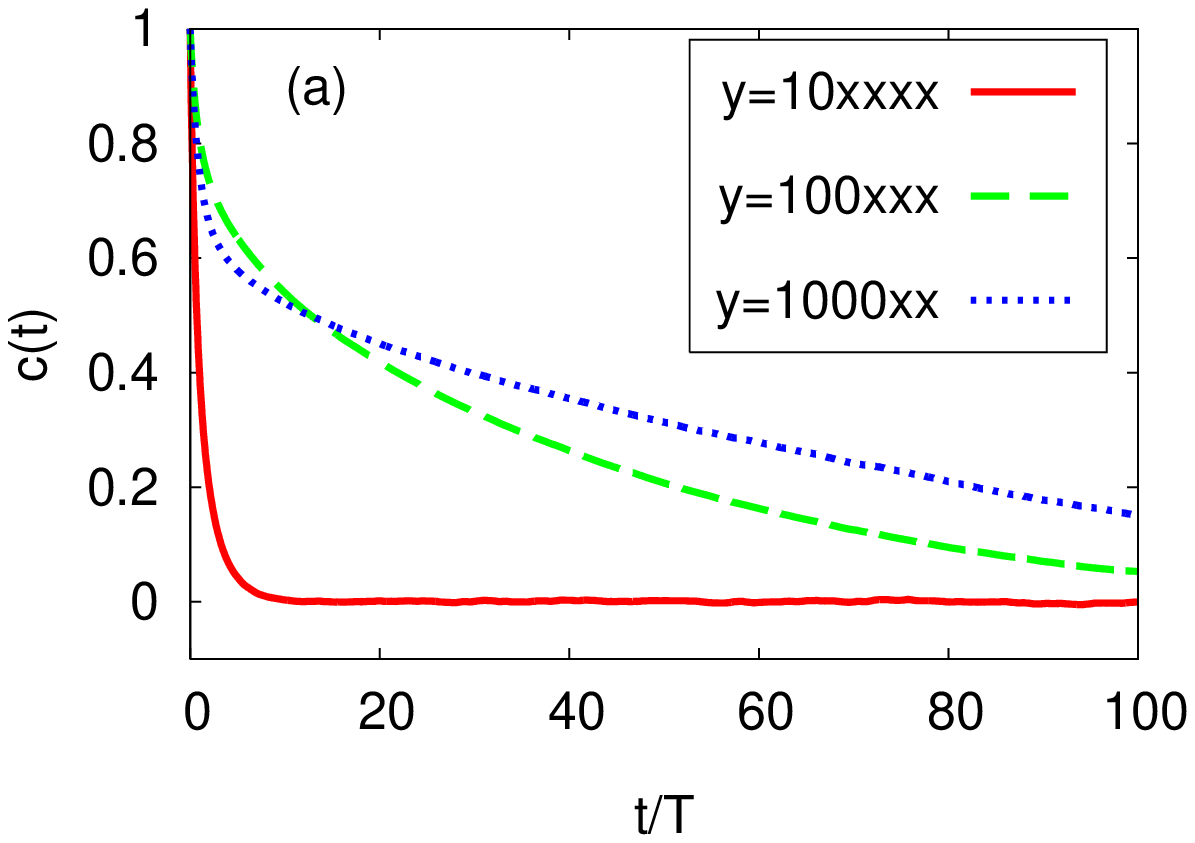}
 \psfrag{t/T}{$\t_c/T$}
 \psfrag{y}{$y$}
 \psfrag{(b)}{$(b)$}
\includegraphics[width=4cm]{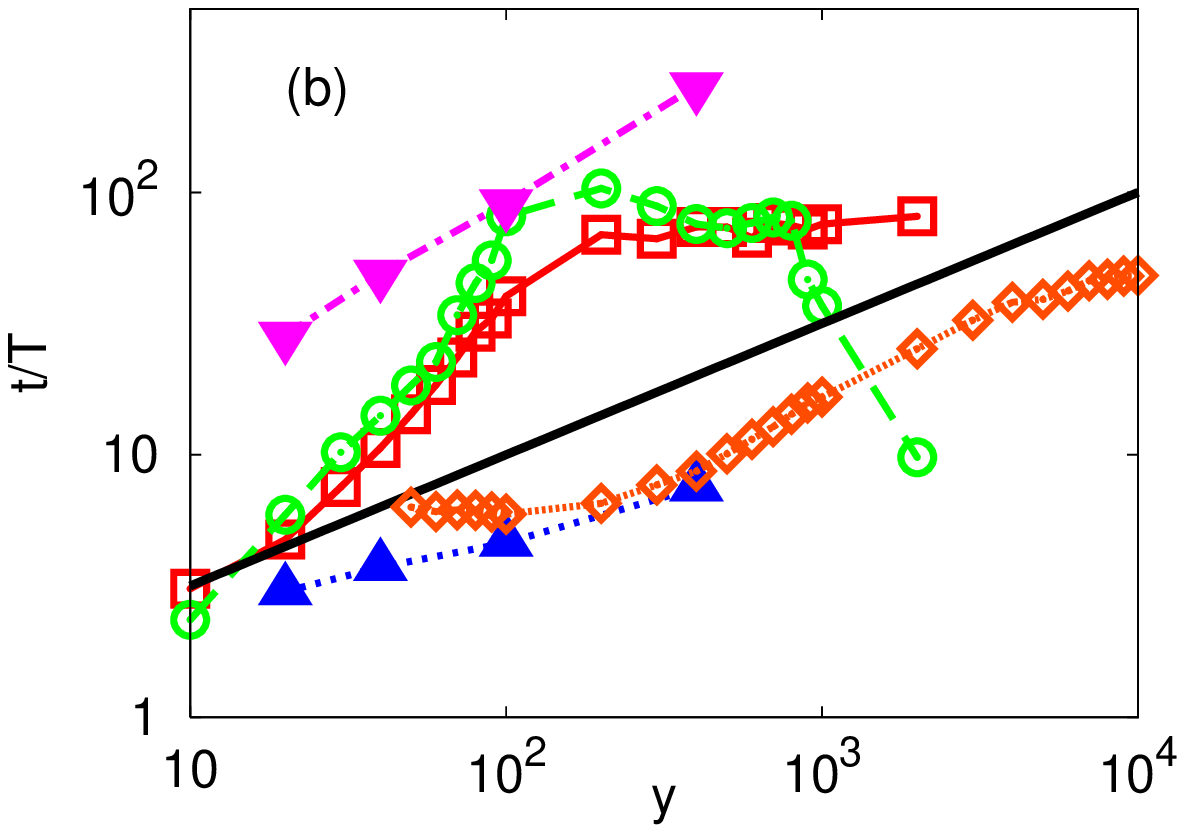}
\caption{(Color online)
($a$) Subtracted auto-correlation $g(t)$ 
 at $y$-levels indicated in the legend for the $L=100$, $N_0=50$ system.
The time series $\bar n(t)$ is collected over $t=200T$ to $2\times 10^4 T$; 
$g(t)$ is further averaged over $30$ initial conditions.
($b$)~The correlation time $\t_c$ ($\Box$) and approach-to-steady-state 
time scale $\t_{ap}$ ($\circ$) as a function of $y$. 
The theoretical time-scales $\t$ (filled $\triangle$), 
$\t_f$ (filled $\nabla$)
(see text) are evaluated from the values of
$a_+,\,b_+,\,c_+$ in Table I.
The solid line is $y^b$ with $b=1/2$.
$\t_c$ ($\Diamond$) for the $L=400$, $N_0=50$ system is also shown.
}
\label{tscale}
\end{figure}

The measured mean fascicle size, obtained by averaging over all the 
existing fascicles at a given $y$  (Fig.~\ref{ap}), grows 
with time as $\bar n = n_\infty - p \exp (-\be t) - q \exp (-t/\t_{ap}) $, 
where  $\t_{ap}(y)$ defines the time scale of 
approach to the steady state value $n_\infty (y)$. 
We find that $\t_{ap}$ 
is an increasing function of $y$ (up to $y \simeq (L/2)^2$ where a 
single fascicle remains and $\t_{ap}$ drops to $T$), and can exceed the 
axon lifetime $T$ by orders of magnitude (Figs.~\ref{ap} and \ref{tscale}$b$). 
Asymptotically in $y$, we find $n_\infty = c+2\r y^b$, with 
$b=0.480\pm0.018$ (Fig.~\ref{ap} inset) -- in reasonable agreement with the 
mean-field prediction. 

The dynamics in the steady state is characterized by the 
auto-correlation function for the mean fascicle size
$\bar n(t)$ at a fixed $y$-level: 
$c(t)=\la \bar n(t) \bar n(0)\ra$ which 
fits to the form $p+q\exp(-\be t)+r\exp(-t/\t_c)$. 
In Fig.~\ref{tscale}$a$ we plot the subtracted correlation function
$g(t)=[c(t)-p]/(q+r)$. The correlation time  $\t_c$ 
increases with $y$ and significantly exceeds the axon lifetime $T$ 
(Fig.~\ref{tscale}$b$). 

%

\section{Effective fascicle dynamics at fixed $y$}
We next examine the dynamics of 
individual fascicles.  These are typically long lived, 
but at a given $y$-level the fascicles very rarely merge 
or split (data not shown). Consequently, the number $n(t)$ 
of axons in each fascicle may be viewed as given by a stochastic 
process specified by the rates $u_\pm(n,y)$ (for transitions 
$n \rightarrow n\pm1$). 
At fixed $y$, a fascicle can loose 
an axon only when the axon dies, thus $u_-(n)=\be n$, 
independent of $y$.

The gain rate $u_+(n,y)$ is governed by 
the properties of the fascicle {\em basin} (see Fig.~\ref{conf}$b$). 
Under the ``always attach, never detach'' rule, new axons 
initiated anywhere within the basin of size $D$ cannot escape the fascicle. 
In addition, some of the axons 
born in the two gaps (of size $E$) flanking the basin contribute. 
Therefore,
\bea
u_+=\a D/L+(\a E/L)[1-\Pi(E,y)]
\label{up}
\eea
where $\Pi(E,y)$ is the probability that 
an axon born within the gap of size $E$ survives as a single axon 
at level $y$. 

\begin{figure}[t]
\psfrag{n}{$n$}
 \psfrag{t/T}{$t/T$}
 \psfrag{n}{{$n$}}
\psfrag{Lb}{{$D$}}
 \psfrag{a}{{$S$}}
 \psfrag{Lb, n}{$D,~n$}
 \psfrag{(a)}{$(a)$}
\includegraphics[width=4cm]{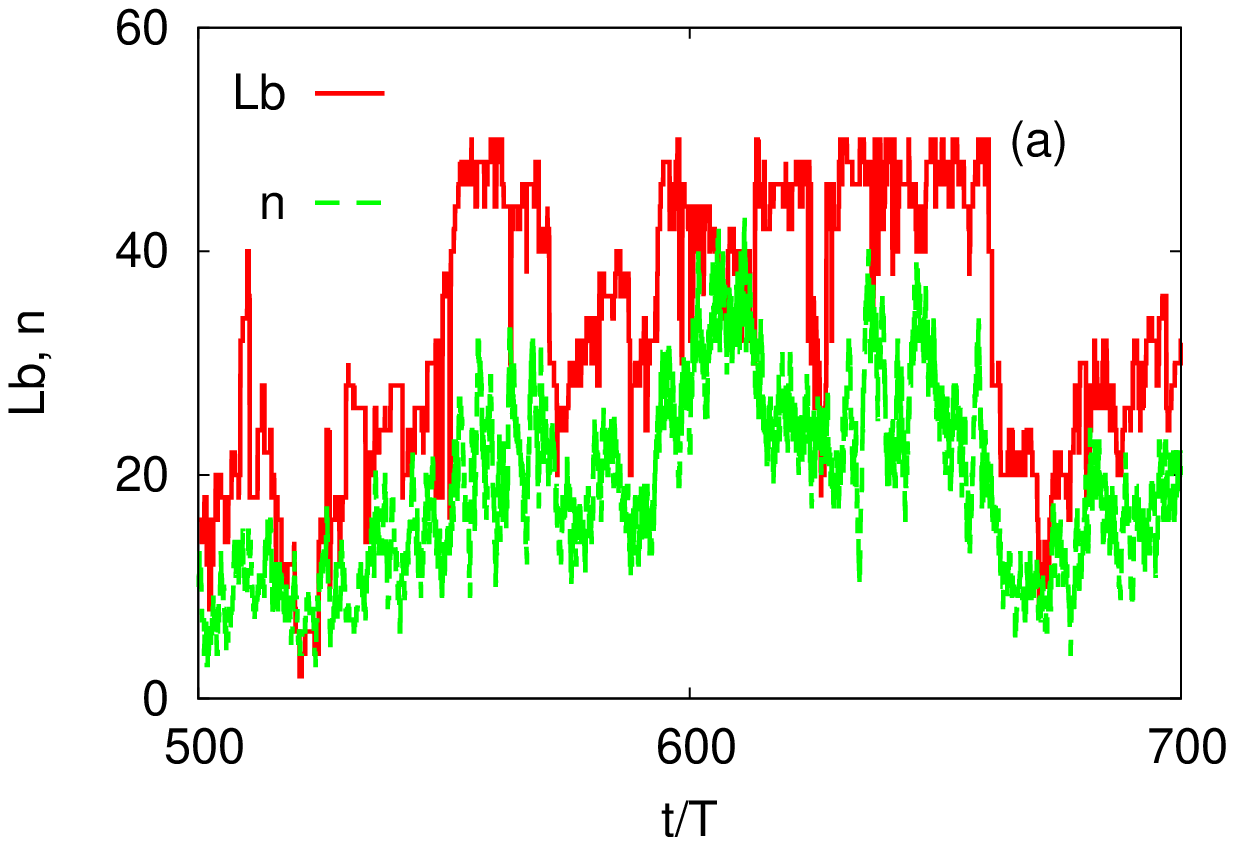}
\psfrag{n}{$n$}
 \psfrag{upbybeta}{$u_+/\be$}
 \psfrag{u1}{$u_+^{(1)}$}
\psfrag{u2}{$u_+^{(2)}$}
 \psfrag{un}{$u_-$}
 \psfrag{up, um}{$u_+, u_-$}
 \psfrag{(b)}{$(b)$}
\includegraphics[width=4cm]{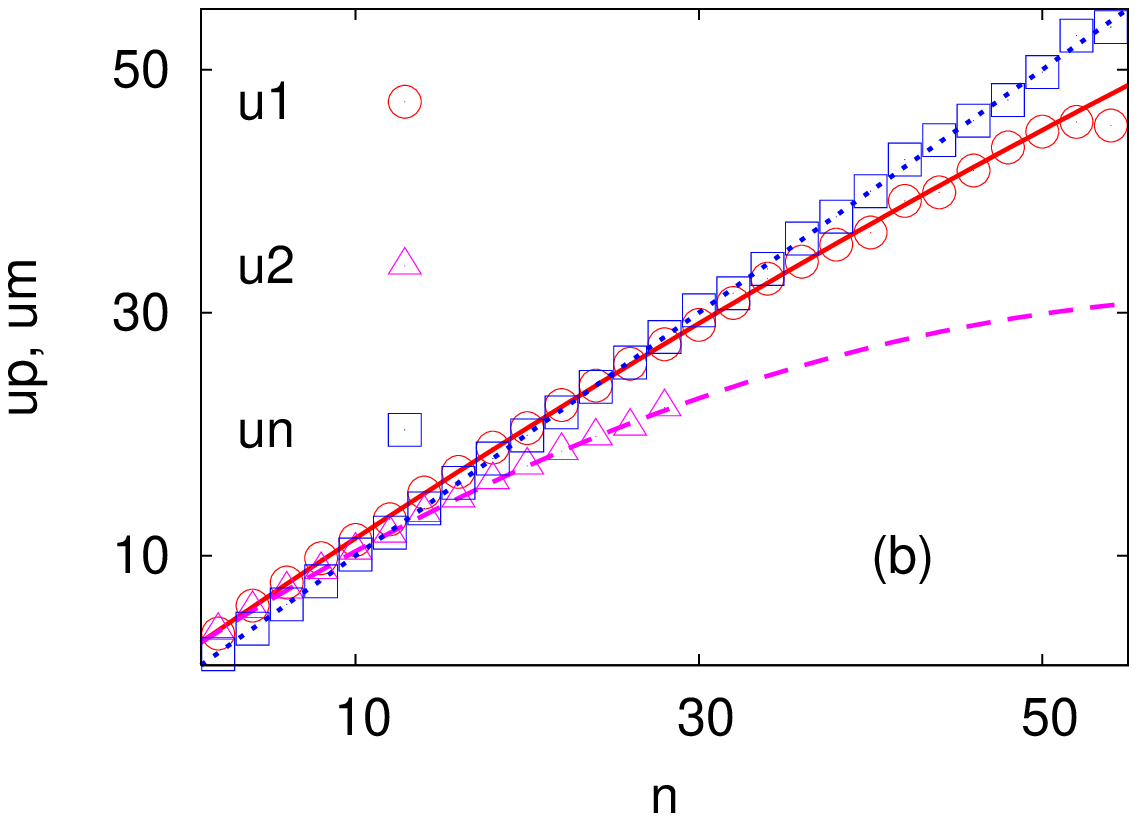}
\caption{(Color online)
($a$)~Time series of number of axons $n$ and basin size $D$ 
for an individual fascicle in a system of
$L=100$ and $N_0=50$ at $y=400$.
The equal time cross correlation coefficient averaged over this
time-span,
$c(D,n)=[\la D n\ra -\la D\ra \la n\ra]/ 
\sqrt{[\la D^2\ra - \la D\ra^2][\la n^2\ra - \la n\ra^2]}=0.74$.
($b$)~Mean gain and loss rates (in units of $\be$) 
[averaged over $10^3$ initial conditions and the interval 
$100T\leq t\leq 150T$]
as a function of fascicle size $n$. The rates are measured at $y=400$ in 
a system with $L=100$. 
The fits (lines) are $u_- =  n$, 
$u_+^{(1)}=1.92+0.974n-0.002n^2$ (for $N_0=50$), 
and $u_+^{(2)}=1.95+0.927n-0.008n^2$ (for $N_0=25$).}
\label{track}
\end{figure}

The two stochastic variables that fully characterize the dynamics of a
fascicle are the  number of axons $n(t)$ and the basin size $D(t)$.
In Fig.~\ref{track}$a$, we plot $n(t)$ and  
$D(t)$ for a specific fascicle followed over $200 T$.  
It is seen that $n$ and $D$ tend to co-vary (cross correlation
coefficient $c(D,n)=0.74$).
In the following we treat the dynamics of $D$ as slave to the 
dynamics of $n$, i.e. we assume that the average separation
 $S = D/(n-1)$ between two neighbouring axons within the basin is 
time-independent. This implies
$u_+(n) = a + b n$
with $b=\be \r S$.

The measured average gain rate $u_+(n)$ in the steady state \cite{rates} 
deviates from linearity at high $n$, but is well fit by 
$u_+(n)=a_+ +b_+ n -c_+n^2$ (Fig.~\ref{track}$b$). The quadratic correction 
may be understood as a saturation effect which reflects that basins of size 
$D$ can not exceed $2 y$ or $L$ and $D > 2 y^{1/2}$ occur with low 
probability.  
The quadratic correction to the linear growth of $u_+$ with $n$ 
becomes significant at $n \gtrsim 2y^{1/2}\r\approx \bar n$. 
Note that the coefficients
$a_+,~b_+$ and $c_+$ are functions of $y$.

{\em Master equation at fixed $y$:}
Let $P(n,t)$ be the distribution of fascicles of size $n$
at time $t$ at a fasciculation distance $y$.
The master equation of the effective birth-death process may be written as 
\bea
\dot P(n,t) &=& u_+(n-1)P(n-1,t)+ u_-(n+1) P(n+1,t)\nn\\ 
&-& [u_+(n)+ u_-(n)] P(n,t),
\label{master}
\eea
for $n>1$. 
For the boundary state ($n=1$)  
$$\dot P(1,t) = J_+(y) + u_-(2) P(2,t) - [u_+(1)+u_-(1)]P(1,t)$$
where $J_+(y)$ represents the rate with which new single 
axons appear between existing fascicles at $y$. 
In the steady state, $J_+(y)$ 
is balanced by the rate with which existing fascicles disappear from the 
system, i.e. $J_+(y)=u_-(1)P_s(1)$. The steady state condition 
$\dot P(n,t)=0$ then implies pairwise balance, 
$u_+(n-1) P_s(n-1) = u_-(n) P_s(n)$, for all $n>1$.  
Thus the steady state distribution is given by 
$P_s(n)=J_+(y) \f{1}{u_-(n)} \prod_{k=1}^{n-1} \f{u_+(k)}{u_-(k)}$,
with the normalization condition $\sum_{n=1}^{\infty} n P_s(n,y)=N$.
In order to obtain a closed-form expression for $P_s(n,y)$, we 
expand the pairwise balance condition up to linear order in $1/N$ 
and solve to
find,
\bea
\be P_s(n,y) \simeq J_+(y)~ n^\g \exp[-\ell (n-1) - \k (n-1)^2],
\label{lin2}
\eea
where $\g=a_+/\be-1$, 
$\ell=1-b_+/\be$ and 
$\k=c_+/2 \be$.

{\em Time scales:}
Three distinct time scales may be extracted from the effective 
birth-death process. 
The correlation time $\t$ for the fascicle size $n$, 
near the macroscopic stationary point $n_s$ [$u_+(n_s)=u_-(n_s)$]
can be expressed\cite{vanKampen} as
$\t=1/(u'_-(n_s)-u'_+(n_s))
=1/(\be - b_+ + 2c_+ n_s)$.
With a linear approximation
of $u_+(n)=a_+ +b_+ n$ the approach-to-steady-state
time scale for $\la n\ra$ is $\t_{ap}=1/(\be - b_+)$ \cite{vanKampen}.
%
We note that the long time scales do not simply arise as a consequence of
fascicles containing many axons, but are due to the dynamics of the
fascicle basins. To see that, imagine a fascicle with frozen boundary 
axons, for which $u_+ \approx (\a/L)D$  with $D$ 
constant.  In this case $u'_+(n_s)=0$ and
the correlation time  $\t\simeq T$.
To obtain $\t >T$, $D$ must co-vary with $n$.
At high $y$, $u'_+\approx 1/T$ resulting in $\t \gg T$.
We note that in the full system, the dynamics of the basin
size can be viewed as arising from the competetion between
neighbouring fascicles for basin space.

A third time scale $\t_f$ can be defined by $J_+(y)=m/\t_f$,
and reflects the rate of turnover of {\em fascicles} in the
full system. Evaluating,
$\t_f=[\int_1^\infty P_s(n,y)dn]/J_+(y)$ using $P_s(n,y)$ 
from Eq.~\ref{lin2} (with $\g=1$), we obtain 
\bea
\t_f=(T/2\k)\left[1-\left(\sqrt\pi e^{\f{\ell^2}{4\k}}(\ell-2\k) \text{erfc}(\ell/2\sqrt\k)\right)/2\sqrt\k\right].
\eea

\begin{figure}[t]
\begin{center}
\psfrag{Ps(n;y)}{$P_s(n,y)$}
 \psfrag{n}{$n$}
 \psfrag{APsny}{$AP_s(n,y)$}
\psfrag{Bn}{$Bn$}
\includegraphics[width=6cm]{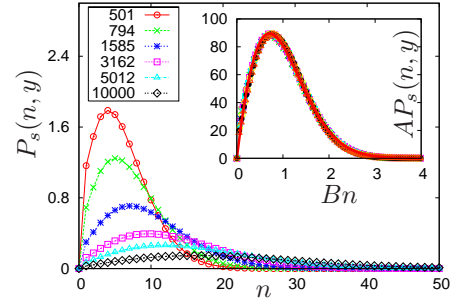}
\end{center}
\caption{(Color online)
Steady state distribution of fascicle sizes $P_s(n,y)$ 
[averaged over $10^4$ initial conditions and the time interval
$10T\leq t\leq 25T$] for the $N_0=100$, $L=800$ system at $y$-levels 
indicated in the legend. 
Inset: A scaling with $B=1/\la n\ra$ and $A=\la n\ra^{2.1}$ collapses all data 
obtained for $y=1585,\, 1995,\, 3162,\, 5012,\, 6310,\, 7943,\, 10^4$ 
onto a single curve
$\phi(u)={\cal N} u \exp(-\nu u-\l u^2)$ 
with $u=n/\la n\ra$ and 
${\cal N}=274$, $\nu=0.78$, $\l=0.45$.  
}
\label{pkg-clps}
\end{figure}

\section{Fascicle size distribution in the steady state} 
Following the discussions on effective single fascicle dynamics,
we now return to the simulation results for the dynamics of the
whole system.
The steady state is characterized by the stationary distribution 
of fascicle sizes $P_s(n,y)$, defined as the number of fascicles of size $n$ 
at level $y$. For a system with $L=800$ and $N_0=100$, $P_s(n,y)$ is shown 
at a series of $y$-levels in Fig.~\ref{pkg-clps}. 
Within the range $y=10^3-10^4$ all data collapse onto a single curve after 
appropriate rescaling (Fig.~\ref{pkg-clps}). 
This data collapse implies the scaling law
\bea
P_s(n,y)= \la n(y)\ra ^{-2.1}\phi(n/\la n (y)\ra)
\label{scale}
\eea
where the scaling function $\phi(u) = {\cal N} u \exp(-\nu u - \l u^2)$.

\section{Analogy to particle aggregation in one dimension}
As we have stressed in the introduction, 
the full dynamics of our system can not be mapped
onto the particle dynamics of a one-dimensional reaction-diffusion system. 
At fixed time $t$ (within the steady state), however,
the aggregation of fascicles with increasing $y$ may be formally viewed as 
the evolution of an irreversible aggregation process 
$mA+nA\to (m+n)A$ in one spatial dimension, where the $y$-coordinate
(Fig.\ref{conf}($b$)) takes the meaning of time. 
This  exhibits analogous scaling properties, but 
with a different scaling function $u\exp(-\l u^2)$ \cite{redner}, which
lacks the exponential part $\exp(-\nu u)$.
It is interesting to note at this point that  
to remove the exponential
part in the expression of the steady state distribution (Eq.\ref{lin2})
one would require $b_+=\be$ which in turn implies
$\t_{ap}=\infty$. In other words, without the exponential part
$\exp(-\nu u)$
the emerging long time scale $\t_{ap}$ is lost. 
This is consistent with the fact that the time scales arising 
from turnover are undefined in the one dimensional analogy.

\section{Scaling of $y$-dependent parameters}
The expansion coefficients for $u_+(n,y)$, measured at selected $y$-levels in a system with $L=100$ and $N_0=50$, are listed 
in Table I. The scaling property of the distribution of fascicle sizes 
(Eq.~\ref{scale}) implies 
$P_s(n,y) 
= {\cal N} y^{-3.1\, b}\, n \exp(-\nu n y^{-b} - \l n^2 y^{-2b})$.
Consistency with Eq.~\ref{lin2} requires $J_+(y)\sim y^{-3.1\, b}$,
$\k\sim y^{-2b}$ [i.e. $c_+\sim y^{-2b}$], 
and  $\ell\sim y^{-b}$ [i.e. $(\be -b_+)\sim y^{-b}$]. 
This is in reasonable agreement with the data in Table I. 
The coefficient $a_+ \simeq 2 \be$ for all $y$ 
reflects $\g\simeq 1$. 
Notice that $J_+(y)\propto \Pi(E,y)$ the probability that an
axon born in the gap remains single at the level $y$.
$\Pi(E,y)$ can be approximated as the survival probability of a 
random walker moving in between two absorbing boundaries (basin borders)
each of which is undergoing random walk; 
thus $\Pi\sim y^{-3/2}$\cite{redner}. This is consistent with the
measured exponent in  $J_+(y)\sim y^{-3.1 b}$.

The gradual approach of $b_+\approx\be \r S$ to $\be$ with increasing $y$
may be understood as follows. At low $y$, fascicles are formed preferentially 
by axons that started with small separation $S$ at the $y=0$ level, and the
typical gap size $E$ exceeds $S$. With increasing $y$, the smallest gaps
are removed to become part of fascicles; consequently both
$E$ and $S$ grow with $y$, and $S$ approaches $1/\r$ at high $y$. The time
averaged $S$ for selected fascicles is shown in the last two columns of
Table I.

\begin{table}[t]
\begin{tabular}{|c|c|c|c|}
\hline
$y$ & $a_+/\be$ & $b_+/\be$ & $c_+/\be$ \\ \hline
20 & 1.96 & 0.862 & 0.0116 \\
40 & 1.94 & 0.909 & 0.0082 \\
100 & 1.96 & 0.947 & 0.0056 \\
400 & 1.92 & 0.974 & 0.0022 \\
\hline
\end{tabular} 
\begin{tabular}{|c|c|}
\hline
$y$ & $\la S\ra$\\ \hline
200 & 1.14\\
400 & 1.60\\
1000 & 2.02\\
10000 & 2.21 \\
\hline
\end{tabular}
\caption{$y$-dependence of the fascicle parameters  defined in the main text.
System parameters are $L=100$ and $N_0=50$.}
\end{table}

{\em $y$-dependence of time scales:}
The $y$-dependence of parameters discussed above implies the
following scaling for the theoretically derived time scales:
the approach-to-steady-state time $\t_{ap}=1/(\be -b_+)\sim y^b$, 
the correlation time $\t\sim\t_{ap}\sim y^b$ and 
the life time of fascicles $\t_f\sim 1/\k\sim y^{2b}$.  
As seen in Fig.~\ref{tscale}$b$, 
the measured correlation time $\t_c$ falls in between the 
computed time scales  $\t$ and $\t_f$, and the three time scales
show distinct $y$-dependence.  
In a system with lower axon density, the time scales are 
reduced (Fig.~\ref{tscale}$b$) as the increased inhomogeneity of
axon separations at $y=0$ leads to a stronger deviation of $\r S$
from $1$.

\section{Conclusion}
To summarize, we have proposed a simple model for axon fasciculation
that shows rich dynamical properties. We identified multiple time scales 
that grow  with the fasciculation distance $y$ and become $\gg T$, 
the average lifetime of individual axons.
The slow time scales do not simply arise as a consequence of
fascicles containing many axons, but are due to the dynamics of the
fascicle basins.
Our theoretical results have wider relevance for existing related 
models (e.g. of  insect pheromone trails \cite{ant1, ant2, pheromone} 
and pedestrian trail formation \cite{human,helbing}), 
in which the slow maturation and turnover 
of the connectivity pattern have not been analyzed in detail.

To conclude, we comment on the applicability of our model to
the biological system which we described in the introduction, 
i.e. the developing mammalian
olfactory system. The model presented in this Letter is readily
generalized to
include multiple axon types, type-specific interactions,
and detachment of axons from fascicles \cite{long_paper}.
This is important as in the olfactory system the nasal epithelium
contains neurons belonging to multiple types, distinguished by the
expressed olfactory-receptor\cite{Mombaerts}, with type specific interaction 
strengths\cite{homotypic1}.
(We note, however, that preparing transgenic mice expressing only a single 
olfactory-receptor has recently become a reality\cite{sakano}).
A further modification of the basic model that might be 
required is the introduction of history-dependent axonal
lifetimes\cite{Paul}. 
A comparatively
simpler task is to relate our model to experimental studies
of axon growth in neuronal cell culture \cite{honig1998, hamlin}.

\acknowledgments
We gratefully acknowledge extensive discussions with Paul Feinstein 
on olfactory development and on the formulation of our model. 


\begin{thebibliography}{10}
\bibitem{rates}
At a given $y$-level, we count the axon arrivals and deaths during an extended
  time period, and record the size of the fascicle that was affected in each
  event. By definition, per unit time $u_+(n,y) P(n,y)$ fascicles of size $n$
  gain an axon, while $u_-(n,y) P(n,y)$ fascicles of size $n$ loose an axon  .
\end{thebibliography}

\end{document}